\documentclass{article}

\usepackage[notes,strict,backend=biber,notetype=endonly,doi=false]{biblatex-chicago} 
\usepackage{xcolor}
\usepackage{setspace}
\usepackage[margin=1in]{geometry}

\usepackage{authblk}

\usepackage{endnotes}

\usepackage{graphicx}	
\usepackage{array}
\usepackage{longtable}
\usepackage{amsmath}
\usepackage{amssymb}
\usepackage{hyperref}

\newcolumntype{L}[1]{>{\raggedright\let\newline\\\arraybackslash\hspace{0pt}}m{#1}}

\begin{document}

\title{Neutrino-based tools for nuclear verification \\ and diplomacy in North Korea}

\author[1]{Rachel Carr}
\affil[1]{Department of Nuclear Science and Engineering, Massachusetts Institute of Technology, 77 Massachusetts Ave., 24-607, Cambridge, Massachusetts  02139, United States; recarr@mit.edu}

\author[2]{Jonathon Coleman}
\affil[2]{Department of Physics, University of Liverpool, Merseyside, United Kingdom}

\author[3]{Mikhail Danilov}
\affil[3]{P. N. Lebedev Physical Institute of RAS, Moscow, Russia}

\author[4]{Giorgio Gratta}
\affil[4]{Physics Department, Stanford University, Stanford, California, United States}

\author[5]{Karsten Heeger}
\affil[5]{Wright Laboratory, Department of Physics, Yale University, New Haven, Connecticut, United States}

\author[6]{Patrick Huber}
\affil[6]{Center for Neutrino Physics, Virginia Tech, Blacksburg, Virginia, United States}

\author[7]{YuenKeung Hor}
\affil[7]{School of Physics, Sun Yat-Sen University, Guangzhou, China}

\author[8]{Takeo Kawasaki}
\affil[8]{Department of Physics, Kitasato University, Sagamihara, Japan}

\author[9]{Soo-Bong Kim}
\affil[9]{Department of Physics, Seoul National University, Seoul, Korea}

\author[10]{Yeongduk Kim}
\affil[10]{Center for Underground Physics, Institute for Basic Science, Daejeon, Korea}

\author[11]{John Learned}
\affil[11]{Department of Physics and Astronomy, University of Hawaii at Manoa, Honolulu, Hawaii, United States}

\author[12]{Manfred Lindner}
\affil[12]{Max-Planck Institute for Nuclear Physics, Heidelberg, Germany}

\author[13]{Kyohei Nakajima}
\affil[13]{Graduate School of Engineering, University of Fukui, Fukui, Japan}

\author[5]{James Nikkel}

\author[14]{Seon-Hee Seo}
\affil[14]{Center for Underground Physics, Institute for Basic Science, Daejeon, Korea}

\author[15]{Fumihiko Suekane}
\affil[15]{Research Center for Neutrino Science, Tohoku University, Sendai, Japan}

\author[16]{Antonin Vacheret}
\affil[16]{Department of Physics, Imperial College London, London, United Kingdom}

\author[7]{Wei Wang}

\author[17]{James Wilhelmi}
\affil[17]{Department of Physics, Temple University, Philadelphia, Pennsylvania, United States}

\author[18]{Liang Zhan}
\affil[18]{Institute of High Energy Physics, Chinese Academy of Sciences, Beijing, China}

\date{\today}

\maketitle

\begin{abstract}

We present neutrino-based options for verifying that the nuclear reactors at North Korea's Yongbyon Nuclear Research Center are no longer operating or that they are operating in an agreed manner, precluding weapons production. Neutrino detectors may be a mutually agreeable complement to traditional verification protocols because they do not require access inside reactor buildings, could be installed collaboratively, and provide persistent and specific observations. At Yongbyon, neutrino detectors could passively verify reactor shutdowns or monitor power levels and plutonium contents, all from outside the reactor buildings. The monitoring options presented here build on recent successes in basic particle physics. Following a dedicated design study, these tools could be deployed in as little as one year at a reasonable cost. In North Korea, cooperative deployment of neutrino detectors could help redirect a limited number of scientists and engineers from military applications to peaceful technical work in an international community. Opportunities for scientific collaboration with South Korea are especially strong. We encourage policymakers to consider collaborative neutrino projects within a broader program of action toward stability and security on the Korean Peninsula.

\end{abstract}

\section*{Context: Shutdown or repurposing of reactors at Yongbyon}

North Korea has built and operated nuclear reactors since the 1960s. As far as public evidence indicates, all functioning reactors have been at the Yongbyon Nuclear Research Center. Plutonium for North Korea's nuclear weapons program has come from a 5\,MW$_{e}$ (20\,MW$_{th}$)\endnote{The distinction between electric power, denoted by a subscript $e$, and thermal power, denoted by a subscript $th$, is important because neutrino emissions are proportional to thermal power. We refer to the 5\,MW$_e$ reactor by that name because it is the more commonly used label.} graphite-moderated, gas-cooled, natural uranium-fueled reactor first operated in 1986 \autocite{puzzle}. Also at Yongbyon is a $100$\,MW$_{th}$ experimental light water reactor (ELWR), fueled with low-enriched uranium (LEU) \autocite{Hecker:2010} and apparently approaching operation \autocite{ELWRNYT}. Yongbyon hosts another small research reactor operated intermittently since the 1960s, remnants of a 50\,MW$_{e}$ reactor project decommissioned in the 1990s, facilities for nuclear fuel fabrication and reprocessing, and a uranium enrichment plant \autocite{puzzle,hecker}.

Leaders within and outside North Korea have proposed the retirement of Yongbyon facilities as a move toward reducing international tensions. The Pyongyang Joint Declaration of September 2018 expresses North Korea's openness to ``permanent dismantlement of the nuclear facilities'' at Yongbyon in exchange for U.S. actions \autocite{pyongyang}. U.S. officials voiced support for complete, verified dismantlement \autocite{pompeo}. An important step in dismantlement would be shutdown of the reactors. This step would precede removal of reactor buildings by months to years to allow residual radioactivity to decay. As an alternative or precursor to full dismantlement, a former U.S. nuclear official has suggested ``demilitarization'' of Yongbyon \autocite{hecker}. Demilitarization could proceed via cooperative conversion of the reactors from weapons preparation to civilian uses such as power generation and medical isotope production.

Whatever goal policymakers pursue for Yongbyon, they will seek concrete, objective assurance that agreed limits are upheld. For reactors, traditional verification protocols involve visual inspections and quantitative assays of fuel and other materials. Standard measurements include weight checks, analysis with gamma and neutron detectors, and sample collection for analysis in off-site laboratories. The International Atomic Energy Agency (IAEA) typically assumes responsibility for these tasks. North Korea has a complicated history with IAEA inspections, and alternative verification methods that require less site access may be desirable. Satellite imaging can often show when reactors produce heat. However, emitted heat is only a coarse indicator of the reactor state and presents little distinction between civilian and military operations. Satellite observation in the visible and infrared cannot penetrate cloud cover and may miss low-power operations.
 
Neutrino detectors could complement traditional verification tools in ways that may appeal to all parties. North Korean officials may be more willing to accept neutrino detectors than standard reactor inspections because neutrino detectors do not require access inside the reactor buildings, which may have other sensitive contents. At the same time, the United States and other parties may value the more persistent and specific information supplied by neutrino detectors, compared to satellite imagery. Both sides may value the opportunity to work together at Yongbyon on a scientifically advanced project with no historical precedent. In this way, neutrino detectors could be a low-stakes step toward more comprehensive inspections at Yongbyon, helping to build trust and lay the groundwork for further cooperative actions. Another option would be to install one or more neutrino detectors at the time of initial on-site inspections. If follow-up inspections are delayed due to a subsequent diplomatic setback, the neutrino detectors could continue to provide monitoring data until on-site inspections resume.

\section*{Technical principle: Reactor monitoring with neutrinos}

Using neutrinos to remotely monitor reactors was first proposed in 1978 by physicists in the Soviet Union~\autocite{Borovoi:1978}. Neutrinos are a byproduct of nuclear fission, arising when neutron-rich fission fragments undergo beta decay. Because they interact only through the weak force, neutrinos from a reactor core pass through the containment building with virtually no attenuation. Roughly $10^{19}$ neutrinos per second flow isotropically from a 100\,MW$_{th}$ reactor. This flux cannot be altered or contained with shielding.

Neutrino emissions carry information directly from the reactor core in real time. Specifically, neutrino emissions bear information about the reactor power level and fuel evolution. The connection between neutrino emission rate and reactor power is simple: both are proportional to the number of fissions occurring in the core. Beyond that proportionality, the neutrino rate is modulated by the mixture of isotopes undergoing fission. In a typical reactor, this mixture contains $^{235}$U, $^{239}$Pu, $^{238}$U and $^{241}$Pu. The plutonium isotopes, including weapons-usable $^{239}$Pu, come from neutron capture on the uranium fuel. The longer a reactor runs, and the higher the reactor power, the more plutonium will be produced. 

Each fissioning isotope produces neutrinos at a different rate. For example, $^{239}$Pu produces about two-thirds as many neutrinos per fission as $^{235}$U. The neutrino energy spectrum also differs between fuel isotopes. For instance, $^{239}$Pu produces a lower-energy neutrino spectrum than $^{235}$U. Observing the number and energy spectrum of neutrinos emitted by a nuclear reactor can therefore reveal the power level and fuel composition of the reactor. Over several weeks of observation, the power history and fuel evolution can be independently constrained without access to operational records~\autocite{Christensen:2015}. 

Physicists have detected neutrinos from reactors for over 60 years. The most accessible detection channel for reactor neutrinos is inverse beta decay (IBD). In this reaction, a neutrino\endnote{Specifically, an antineutrino of the electron flavor. This is the only type of neutrino produced in reactors.} interacts with a hydrogen nucleus, yielding a positron and neutron. Proton-rich targets, such as water and hydrocarbons, make ideal detector media. Over decades of neutrino detector evolution, organic scintillators have remained the medium of choice for detecting IBD because of their good energy resolution and moderate cost. Ongoing R\&D may yield other techniques for observing neutrinos at reactors \autocite{conus}. Here, we focus on IBD in scintillators as an available, well-demonstrated option.

The world's first observation of neutrinos occurred at a plutonium production reactor at the U.S. Atomic Energy Commission's Savannah River site in the 1950s~\autocite{Cowan:1992xc}. In the early 2000s, the much larger KamLAND experiment measured neutrino flavor oscillations from power reactors in Japan, key evidence in establishing that neutrinos have mass~\autocite{Eguchi:2002dm}. In the mid-2010s, precision neutrino measurements occurred at reactors in China~\autocite{An:2012eh}, South Korea~\autocite{Ahn:2012nd}, and France~\autocite{Abe:2011fz}. Recently, searches for sterile neutrinos, a hypothetical particle beyond the standard model of particle physics, have spurred the development of high-precision, surface-deployable detectors. These compact, relatively simple detector designs also happen to be ideally suited for reactor monitoring.

At present, hundreds of neutrinos are detected daily from commercial and research reactors in East Asia, Europe, and the United States. Over five million reactor neutrino interactions have been recorded and analyzed to date. Using neutrino data to observe reactor power levels and fuel evolution is now common in particle physics experiments, as one step in more complex analyses~\autocite{An:2017osx,RENO:2018pwo}. Since at least the early 2000s, national and international agencies have recognized the potential to apply this technology to practical problems \autocite{iaea}.

In the following sections, we present options for using neutrino detectors as verification tools at Yongbyon. We outline three specific deployment scenarios. The first option is using neutrino detectors to verify that the 5\,MW$_{e}$ reactor, ELWR, or both are shut down. The second is using a neutrino detector to verify that the ELWR is running for the civilian purpose of electricity generation and not for weapons production. Each of these two options could be realized near the reactor buildings, using demonstrated technology, within about one year following development of a specific deployment plan. A third option is a larger neutrino detector which could verify shutdown of both reactors from a distance of up to 1 km.

\begin{figure*}
    \includegraphics[width=\textwidth]{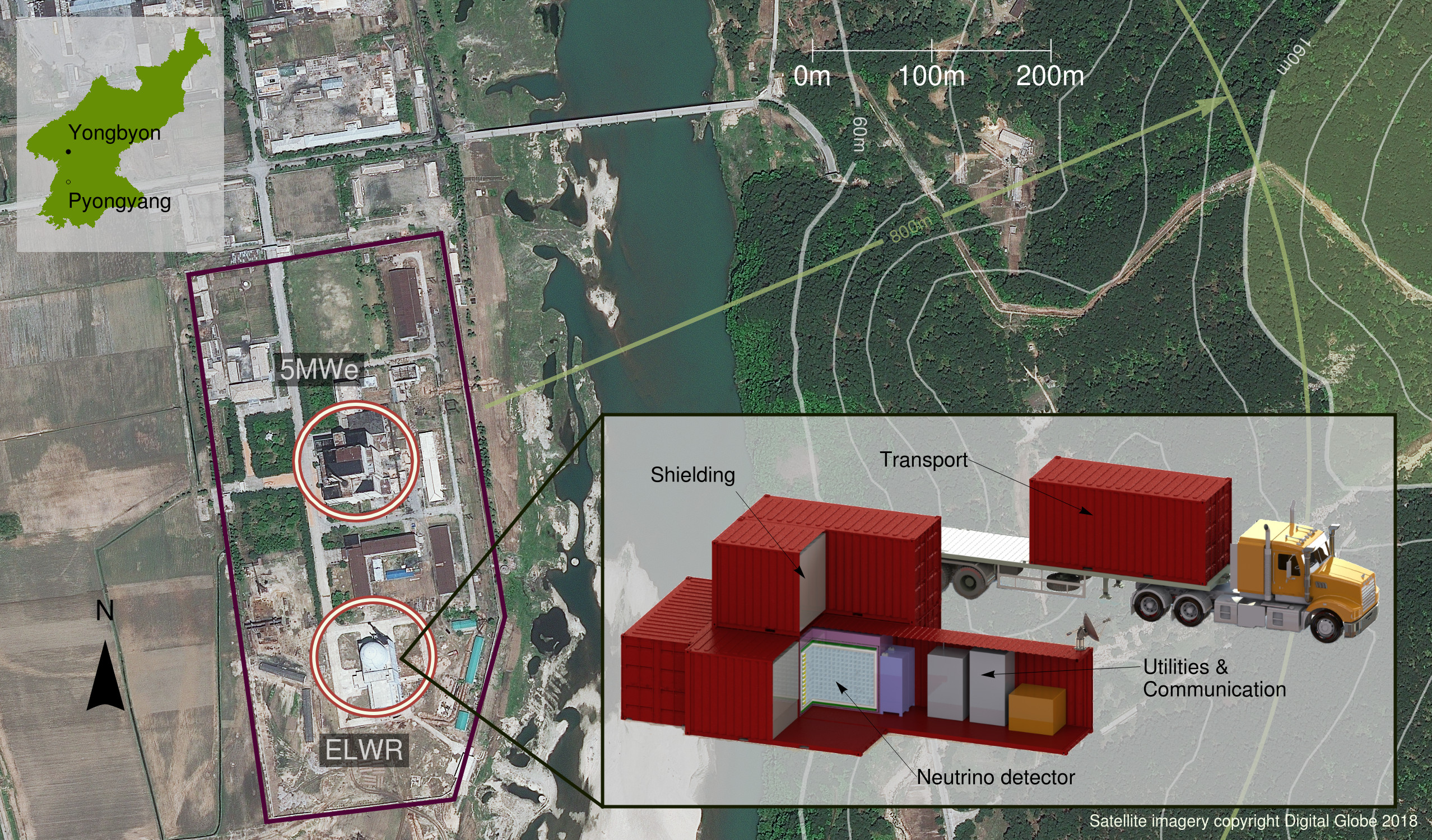}
    \caption{\label{fig:container} Opportunities for neutrino-based verification are highlighted on a satellite image of part of the Yongyon site; see text for full explanation. Satellite image copyright DigitalGlobe (2018).}
\end{figure*}

A possible deployment scenario is sketched in Figure \ref{fig:container}. The red and white circles around the 5\,MW$_e$ reactor and ELWR correspond to a radius of roughly 50\,m. Locations suitable for cooperative deployment of neutrino detectors appear as near as 20\,m from each core, as noted in a previous analysis \autocite{Christensen:2014pva}. The purple enclosure indicates a possibly fenced area; a detector could be deployed outside at a standoff of slightly over 100\,m. The inset at lower right is a concept for a detector and shielding scheme housed in ISO freight containers, along with a possible mechanism for transporting the detector to the site. The shaded region at top right indicates where a larger detector in a horizontal tunnel could have an overburden of at least 100\,m, at a standoff of about 800\,m (indicated the arrow and arc). 

\section*{Neutrino-based verification of reactor shutdowns at close range}
\label{sec:nearonoff}

If the reactors at Yongbyon shut down, neutrino detectors could verify that they remain off during the cool-down period of months to years preceding full dismantling. To do this, detectors would watch for anomalous neutrino emission from the cores. Shutdown reactors and spent fuel emit a very low-level neutrino flux due to the decay of long-lived fission products. This reactor-off neutrino flux typically decays below detection threshold a few days after shutdown. The neutrino flux from a reactor operating at even low power levels is much higher than the reactor-off flux. Thus, restart of a reactor registers as a distinct signal in a suitably sensitive neutrino detector.

Reactor on-off transitions have been observed in neutrino detectors since the 1980s. Physicists in the Soviet Union pioneered these measurements at the Rovno reactor \autocite{Afonin:1985rw}. U.S. physicists explored the idea in the early 2000s \autocite{Bernstein:2001cz} and performed a similar experiment at the San Onofre Nuclear Generating Station \autocite{Bowden:2006hu}. Further demonstrations of shutdown/restart observation followed in an applications-oriented experiment in France \autocite{Boireau:2015dda}, as well as basic physics experiments in France \autocite{Abe:2011fz}, China \autocite{An:2012eh}, and South Korea \autocite{Ahn:2012nd}. One lesson from this progression is that detector segmentation is key to background rejection, allowing detectors to operate on the earth's surface with essentially no cosmic ray shielding. 

Today, a state-of-the-art detector can observe an off-to-on (or on-to-off) reactor transition within hours, depending on the detector size and reactor proximity. A notable example is the PROSPECT detector at Oak Ridge National Laboratory, operating with a total mass of 4 tons and less than 1 meter-water-equivalent overburden~\autocite{Ashenfelter:2018iov}. In the last two years, the PROSPECT (United States), NEOS (South Korea) \autocite{Ko:2016owz}, DANSS (Russia) \autocite{Alekseev}, CHANDLER~\autocite{Haghighat:2018mve}, Neutrino-4 (Russia) \autocite{Serebrov}, STEREO (France) \autocite{Almazan:2018wln}, and SoLi$\delta$ (Belgium) \autocite{solid} experiments, as well as a detector at the Wylfa reactor (UK) \autocite{coleman}, have observed differences in neutrino event rates between reactor-on and reactor-off periods. Some of these experiments have demonstrated steps toward field readiness, including SoLi$\delta$ and the Wylfa detector (now upgraded to the VIDARR project), both deployed in ISO freight containers, and CHANDLER, deployed in a road-mobile trailer. Collaborations are also pursuing IBD signals from reactors in Brazil (the Angra detector, collecting data at the power reactor of the same name)\autocite{Anjos:2015wxj}, Japan (the PANDA experiment, anticipating a deployment at the Ohi power reactor)\autocite{Kuroda:2012dw}, and India (the ISMRAN detector under development at the Bhabha Atomic Research Center)\autocite{Mulmule:2018efw}. 

Detectors using similar technology could be deployed at Yongbyon. These could verify continuous shutdown of the 5\,MW$_{e}$ reactor, ELWR, or both. At either reactor, the closest conceivable distance for a detector deployment is about 20\,m from the core, as noted in a previous analysis \autocite{Christensen:2015}. This position would be, especially in the case of the 5 MW$_e$ reactor, directly outside the reactor building. We envision the detector being installed outside a reactor which is initially known by all parties to be in the off state and which remains in the off state for long enough (e.g., a few weeks) for the background rate to be measured. The detector would then look for an anomalous rise in the data rate. At the ELWR, a 4-ton segmented scintillator detector could identify an unauthorized start-up of the reactor in at 99\% or greater confidence level within one day in 95\% of cases (with the variation arising from statistical variation in the event counts). At the smaller 5\,MW$_{e}$, a start-up could be detected at 95\% or greater confidence level within two weeks in 95\% of cases. 

These estimates use the measured signal efficiency and background rates of the PROSPECT detector operating at Oak Ridge and a basic rate-based hypothesis test, as described in the Appendix. In both cases, the false positive rate from the simple event-count criterion would be about one per year. Examining the energy spectrum of the events, which differs substantially between background and signal, could eliminate most false positives. In a real monitoring campaign, spectrum shape and time series information could be included in the hypothesis test itself. This would likely increase sensitivity beyond the simple estimates here. The cost of PROSPECT was \$5~million, and the detector was constructed in less than one year.

\section*{Neutrino-based verification of the reactor core state}
\label{sec:core}

As an alternative to a total shutdown of Yongbyon, political leaders may agree to continue operating one or more reactors there for civilian purposes. For example, they could choose to move forward with operating the ELWR. This reactor was designed to generate electricity and is not optimal for producing weapons-usable plutonium~\autocite{Hecker:2010}. However, experts have noted that a modified ELWR could use a different fuel loading and power profile to enhance plutonium production~\autocite{Albright:2015}.

To verify civilian operations, neutrino detectors could observe both the power profile and fuel evolution of the ELWR. As noted in the previous section, neutrino-based tracking of reactor power profiles has been demonstrated in multiple experiments. Neutrino-based tracking of fuel evolution has also been demonstrated. In particular, neutrino detectors at LEU-fueled light water reactors have observed the characteristic change from $^{235}$U-dominated fissions to a mixture of $^{235}$U and bred-in $^{239}$Pu fissions \autocite{An:2017osx, RENO:2018pwo}. These observations used both neutrino rate and spectral shape information. Using similar techniques, simulations show that neutrino detectors can distinguish normal, electricity-producing LWR operations from operations designed to produce weapons-suitable plutonium ~\autocite{Christensen:2014pva, Christensen:2015}.

As a specific example, a neutrino detector outside the Yongbyon ELWR building could check whether the reactor is using a normal, semi-recycled core or has substituted a fresh core and possibly diverted irradiated fuel for weapons. A 20-ton detector using existing scintillator technology could identify the diversion of a core containing 8\,kg of $^{239}$Pu (one significant quantity by IAEA definition) in about 200 days. This estimate assumes the signal efficiency and background levels measured in the PROSPECT detector, as described in the Appendix. An advantage of neutrino-based core monitoring, compared to other plutonium inventory approaches, is that it is possible to reconstruct the plutonium content of a reactor even after a pause in data-taking \autocite{Christensen:2015}. In this way, a neutrino detector could help to recover from a gap in verification data. Detecting plutonium diversion from the 5\,MW$_{e}$ reactor would take longer, likely beyond a useful timeframe, because of this reactor's lower power and because the fission profile evolves less in a graphite reactor than an LWR. 

We emphasize that both the shutdown verification option presented in the preceding section and core monitoring option in this section are achievable with relatively small, surface-deployable systems employing demonstrated technology. The PROSPECT-like detector suggested in the previous paragraph could be assembled off-site, inside a standard shipping container, with lead- and water-filled containers providing adequate cosmic ray shielding. The inset in Figure \ref{fig:container} depicts this concept. On-site infrastructure requirements and data handling needs would be comparable to that of conventional radiation detection systems. 
 
\section*{Neutrino-based verification of reactors over a wider area}
\label{sec:midfield}

So far, we have described options for deploying neutrino detectors within sight of the reactor buildings. These are attractive options because they allow the detector to remain small and relatively simple to construct. With the strong caveat that required detector size (or observation time) scales as the square of the standoff distance, neutrino signals can be detected from farther away.\endnote{Reactor neutrinos have been observed from distances exceeding 100\,km in the world's largest operating liquid scintillator detector, the 1-kiloton KamLAND located 1 km underground~(\autocite{Eguchi:2002dm}). Note that this detector is roughly 30 times larger than the largest option we consider for Yongbyon. Over long observation times, faint reactor neutrino signals have been detected from as far as 1000 km in the 300-ton, very low-background BOREXINO detector located 1.4 km underground~(\autocite{Bellini:2010hy}).} Crucial to these observations are very low background rates. The low background is achieved by locating the detectors underground.

At Yongbyon, it could be feasible to monitor shutdown of both the 5\,MW$_e$ reactor and ELWR from a distance of 800--1000 m. This scenario would require a larger detector than the cases discussed in the previous two sections. A well-demonstrated option is a liquid scintillator detector like those used in the Daya Bay experiment (site of eight such detectors)~\autocite{An:2012eh}, RENO (two detectors)~\autocite{Ahn:2012nd}, and Double Chooz (two detectors)~\autocite{Abe:2011fz}. These detectors require sizable overburden for cosmic ray shielding. The 480\,m-high Yaksan mountain, across the Kuryong river from the Yongbyon reactors, could provide cosmic ray shielding similar to that of Daya Bay, RENO, and Double Chooz. 

In a horizontal tunnel in Yaksan, we estimate that a roughly 30-ton liquid scintillator detector could, in 95\% of cases, detect a change of reactor state from on to off at 99\% or greater confidence level within 15 days for the ELWR. Startup of the 5\,MW$_e$ reactor could be detected at 95\% or greater confidence level in at 95\% of cases within approximately 250 days. The time to make 8\,kg of plutonium (one significant quantity) in the 5\,MW$_e$ reactor is about 400 days. A 250-day warning could be timely by this standard. A roughly 25\% larger detector would be needed to meet the more stringent standard of detecting a reactor startup within the time to produce 4 kg of plutonium (about 200 days), which has been estimated as sufficient for a nuclear weapon \autocite{fetter}. In addition to looking for unauthorized startup of the 5\,MW$_e$ reactor and ELWR, this type of detector could provide an upper limit on all reactor operations within a radius of 1--2 km. The precise size and location of the detector could be tailored to suit the specific monitoring goal. Construction time and cost would be greater than for the options in the previous two sections.

\section*{Options for cooperative neutrino science on the Korean Peninsula}

Neutrino-based verification could be part of a broader set of actions reintegrating North Korea into the international community. If carried out cooperatively, neutrino projects could complement wider efforts to redirect scientists and engineers from the weapons program to peaceful technical work. As we have noted, neutrino physics as an experimental science originated at a military reactor site with a team of weapons physicists \autocite{reinesnobel}. Workforce reengagement was later a key part of cooperative threat reduction programs in the former Soviet republics \autocite{crs}. For North Korea, policy experts have again stressed the value of scientific cooperation to build trust, secure hazardous materials, and help stem the spread of nuclear weapons expertise to other parties \autocite{nunnlugar}. A team of a few dozen scientists and engineers could support a neutrino project at Yongbyon, split between North Korean and foreign personnel. Like other general-purpose detectors, such as Geiger counters, neutrino detectors could be built and operated cooperatively without exchange of classified or weapons-related information. Descriptions of the relevant technology and analyses already appear in publicly available scientific literature, as we have cited in this document. 

Neutrino projects offer a special opportunity to strengthen North-South Korean interactions. South Korea hosts one of the world's major reactor neutrino experiments, the Reactor Neutrino Oscillation Experiment (RENO)~\autocite{Ahn:2012nd}, as well as the ongoing Neutrino Experiment for Oscillation at Short baseline (NEOS)~\autocite{Ko:2016owz} and Advanced Molybdenum-based Rare process Experiment (AMoRE)~\autocite{Bhang:2012gn}. Physicists from South Korea collaborate extensively on projects beyond their borders, including the upcoming Hyper-Kamiokande experiment based in Japan and possibly in Korea \autocite{Abe:2018uyc, Abe:2016ero}. China is also making major new investments in neutrino physics. All of these ventures are pushing limits in electronics design and computing algorithms. For North Korea, participating in international physics collaborations could open the door to valuable scientific and economic opportunities in and beyond the region.

As a first step, policymakers could agree to involve scientists and engineers from North Korea in neutrino-based verification projects at Yongbyon. Beyond that, universities and laboratories outside North Korea could consider student exchanges and visiting professorships in neutrino physics and related areas. Pyongyang's recently completed Sci-Tech complex could host an international particle physics conference. To further North-South unity, neutrino detectors at Yongbyon could be twinned with detectors at power reactors in South Korea. This joint program could explore topics in both basic and applied science. On a small scale, a joint North-South particle physics venture brings to mind the 1954 founding of CERN, one of the first diplomatic agreements between France, Germany, and neighboring nations following World War II.

In closing, we emphasize that technology and expertise are ready to implement any of the options presented in this document. Preparation of a detailed construction plan and cost estimate for one or more specific deployment options could begin immediately. We encourage policymakers to consider neutrino-based options as part of the broader pursuit of stability and security on the Korean Peninsula.

\newpage
\begin{appendix}
\section*{Appendix: Basis of estimates}

The sensitivity estimates for cases with 20 m baselines come from scaling the observed signal and background rates of the PROSPECT detector~\autocite{Ashenfelter:2018iov}. This detector observed 771 signal events per day in 2 tons of fiducial volume (4 tons total volume). The PROSPECT
signal-to-background ratio for IBD-like events is 0.83. The detector is located at a standoff of 7.9\,m from an 85\,MW$_{th}$ reactor. Signal and background rates for different standoffs and reactor powers follow these simple scaling relations:
\begin{eqnarray}
 S&=&771
 \left(\frac{m}{2\,\text{[ton]}}\right)\left(\frac{P}{85\,\text{[MW$_{th}$]}}\right)\left(\frac{7.9\,\text{[m]}}{L}\right)^2
 \,\mathrm{d}^{-1}\,;\label{signal}\\ B&=&\frac{771}{0.832}
 \left(\frac{m}{2\,\text{[ton]}}\right) \,\mathrm{d}^{-1}\,.\label{background}
\end{eqnarray}
When scaling to detector sizes other than the actual PROSPECT size, we assume that the fiducial volume is all but the outer, 15-cm-thick layer of the total scintillator volume. In this scaling, a 12-ton fiducial mass correspond to roughly 20 tons of total scintillator mass.

The sensitivity estimates for the more distantly deployed detector come from scalings similar to Eqs. \ref{signal}-\ref{background}. In this case, the reference detector is a Daya Bay near  detector in Experimental Hall 1 rather than PROSPECT \autocite{An:2012eh}. The Daya Bay detectors in that location have about 250\,m water equivalent overburden, which corresponds to about 100\,m of actual rock overburden.  The Daya Bay detectors have 20 tons of fiducial volume and obtain a signal rate of about 700 events per day and a total background rate of about 12 events per day. The two closest reactors have combined thermal power of 5.8\,GW$_{th}$, and the standoff is about 400\,m. The peak of Yaksan has an elevation of 480\,m and is about 2\,km from both the 5\,MW$_e$ and the ELWR. Locations with 100\,m of rock overburden can be found starting at a distance of about 800\,m from the 5\,MW$_e$ reactor, as shown in Figure \ref{fig:container}.

A simple sensitivity metric is the time $T$ required to detect a transition between reactor-off to reactor-on states at 95\% confidence level (CL) or greater, in at least 95\% of cases. Using simple counting statistics, the criterion for detecting at 95\% CL a transition from a known background rate $B$ to the signal plus background rate $S+B$ in time $t$ is:
\begin{equation}
  \int_{-\infty}^{(S+B)t} dx \, f(x | \mu = Bt, \sigma^2 = Bt) = 0.95
  \label{eq:criterion}
\end{equation}
where $f(x|\mu,\sigma^2)$ is the normal distribution with mean $\mu$ and variance $\sigma^2$. This criterion is met or exceeded in 95\% of cases if the mean expected number of events, equal to $(S+B)T$, satisfies:
\begin{equation}
    \int_{(S+B)t}^\infty dx \, f(x | \mu = (S+B)T, \sigma^2 = (S+B)T) = 0.95
    \label{eq:cases}
\end{equation}
We use Equation \ref{signal}-\ref{eq:cases} to estimate the time needed to detect a reactor-off to reactor-on transition, increasing the standard in Equation \ref{eq:criterion} to 0.99 where needed to reduce the false positive rate.

For the core state analyses, the reactor core simulation for the ELWR is based on the light-water converted IR-40 reactor at Arak, Iran, scaled to a reactor power of 100\,MW$_{th}$~\autocite{Willig:2012}. The reactor core simulation for the 5\,MW$_e$ is from a previous analysis of that reactor~\autocite{Christensen:2015}. The time $t_\mathrm{SQ}$ to produce 8\,kg of plutonium (1 significant quantity, or SQ, by IAEA definition) is 450\,d for the 5\,MW$_e$ and 330\,d for the ELWR. 
The spectral analysis techniques are described in a previous work~\autocite{Christensen:2014pva}.  Note that for the 5\,MW$_e$, a core swap cannot be detected even in a zero-background scenario in less than 500 days, which exceeds the time to make 8\,kg plutonium in this reactor.

%

\end{appendix}

\theendnotes  

\end{document}